\documentstyle[12pt]{article}

\newcommand{\asize}[1]{\renewcommand{\arraystretch}{#1}}

\renewcommand{\theequation}{\arabic{section}.\arabic{equation}}

\newcommand{\be}{\begin{equation}}
\newcommand{\ee}{\end{equation}}
\newcommand{\ba}{\begin{array}}
\newcommand{\ea}{\end{array}}
\newcommand{\dst}{\displaystyle}
\newcommand{\tst}{\textstyle}
\newcommand{\sst}{\scriptstyle}

\newcommand{\bac}{\begin{array}{c}}
\newcommand{\bal}{\begin{array}{l}}
\newcommand{\baR}{\begin{array}{r}}
\newcommand{\bacc}{\begin{array}{cc}}
\newcommand{\ball}{\begin{array}{ll}}
\newcommand{\baccc}{\begin{array}{ccc}}
\newcommand{\barcl}{\begin{array}{rcl}}
\newcommand{\bacl}{\begin{array}{cl}}
\newcommand{\bacll}{\begin{array}{cll}}
\newcommand{\eac}{\end{array}}
\newcommand{\ber}{\begin{eqnarray}}
\newcommand{\eer}{\end{eqnarray}}

\newcommand{\map}[1]{\stackrel{#1}{{\dst \longrightarrow}}}

\newcommand{\dbar}{{\overline{\partial}}}

\newcommand{\Rbar}{{\overline{R}}}
\newcommand{\Sbar}{{\overline{S}}}

\newcommand{\etatil}{{\tilde{\eta}}}
\newcommand{\Atil}{{\tilde{A}}}
\newcommand{\Ctil}{{\tilde{C}}}

\newcommand{\Ptil}{{\tilde{P}}}

\newcommand{\ntil}{{\tilde{n}}}

\renewcommand{\d}{{{\partial}}}
\newcommand{\half}{\frac{1}{2}}

\newcommand{\ra}{\rightarrow}

\newcommand{\DD}{{\cal D}}

\newcommand{\OO}{{\cal O}}

\newcommand{\AN}{-\frac{\lambda A}{2N}}
\newcommand{\ANN}{\frac{\lambda A}{~2N^2}}
\newcommand{\gst}{\left( \frac{-\lambda A}{2N} \right)}
\newcommand{\gsu}{\left( \frac{\lambda A}{2N^2} \right)}
\newcommand{\e}{E_2}

\newcommand{\eb}{\left[ E_2 \right]}
\newcommand{\ep}{E_2^{\, \prime }}
\newcommand{\epb}{\left[ E_2^{\, \prime } \right]}
\newcommand{\epp}{E_2^{\, \prime \prime }}
\newcommand{\eppb}{\left[ E_2^{\, \prime \prime } \right]}
\newcommand{\txthalf}{{\textstyle \half}}
\newcommand{\fp}{F^{+ \prime}}
\newcommand{\fpp}{F^{+ \prime \prime}}
\newcommand{\fpb}{F^{+(4)}}
\newcommand{\fpc}{F^{+(6)}}
\newcommand{\fpd}{F^{+(8)}}
\newcommand{\fpbb}{F^{+(3)}}

\asize{1.8}

\begin{document}

\thispagestyle{empty}
\setcounter{page}{0}
\begin{flushright}
RU-94-58 \\
July, 1994 \\
{\tt hep-th/9407176}
\end{flushright}

\begin{center}
 ~ \\
 ~ \\
 ~ \\
{\Large The String Partition Function} \\
{\Large for QCD on the Torus} \\
 ~ \\
 ~ \\
 ~ \\
{\large \sc Robert E. Rudd}$^{\dag}$ \\
 ~ \\
{\em Department of Physics and Astronomy} \\
{\em Rutgers University} \\
{\em Piscataway, New Jersey~~08855-0849~~USA}
\end{center}

\vspace{.8in}
\centerline{\large ABSTRACT}
\centerline{~}

We study the free energy of the pure glue QCD string with a
torus target space and the gauge groups $SU(N)$ and (chiral)
$U(N)$.  It is highly constrained by a strong/weak
gauge coupling duality which results in modular covariance.
The string free energy is computed exactly
in terms of modular forms for worldsheet genera 1 - 8.
It has a surprisingly mild singularity in the weak gauge
coupling/small area limit.

\vfill

\vspace{.3in}
\footnoterule
$^{\dag}$ \parbox[t]{6in}{
{\small E-mail: ~~rerudd@physics.rutgers.edu} }

\newpage

\section{Introduction}

  The idea of a string theoretic formulation of QCD is as tantalizing today
as it was twenty years ago.  Despite its age and elusiveness, the promise
of a description of the phenomena of strongly coupled gauge theory in
term of strings is compelling.  The many problems encountered in trying
to implement this idea have shown that the formulation must differ
substantially from the critical string.  It must not include spacetime
gravity, for instance.  As a result, there are no promising models at
this time.

  The hope for a QCD string has been rejuvenated recently by significant
progress understanding QCD in two dimensions.  Pure glue QCD$_2$ has been
solved exactly \cite{MR}, and a great deal of work has gone into
constructing a string theory from its large $N$ expansion
\cite{Gross,Min,GT}.  The string action that has resulted is a perturbed
topological sigma model coupled to topological gravity
\cite{Vafa,DR,Hora,GMtop}.  Perhaps a simpler formulation of the
topological string is
possible, but we are in a novel situation of having a string theory
that has been shown to be equivalent to QCD, albeit in two dimensions.

  The ultimate goal is to use the string theory of QCD$_2$ to
construct, directly or indirectly, a QCD$_4$ string action.  The
direct extension of the sigma model to four dimensions would presumably
yield some version of topological Yang-Mills, although this is a
non-trivial, open problem.  A further perturbation would be
necessary to get dynamical (pure glue) QCD.

  If the direct approach proves to be intractable, we may still make
progress toward a QCD$_4$ string by identifying properties of the
string that are independent of dimension.  To that end, we will
examine the QCD string partition function in detail, particularly
focusing on QCD on the torus which most resembles critical string
theory.  The large $N$ expansion of the QCD free energy has been
given in terms of two different group theoretic sums \cite{MR,GT},
the free energy for free fermions and the equivalent Jevicki-Sakita
bosons \cite{Douglas,MP}
and, of course, the free energy of the topological string
itself \cite{GMtop}.  The various formulations readily reveal different
aspects of the free energy, but none is completely explicit.  We
will show how the structure of the Jevicki-Sakita expressions
and the simple algebra of the heat kernel sum combine to let us
compute the free energy efficiently.  The free energy will be
calculated exactly in terms of modular forms up to genus 8.

  It is very rare to have an exact expression to the eighth order
of string perturbation theory.  In those cases where such an
expression has been found, as in matrix models, it is possible
to continue to all orders.  In fact, there is a differential
equation relating the free energy at a given order to that at
lower orders.  If such an equation were known for the QCD string,
then there would be no need to display the horrendous expression
for genus 8.  But the obvious candidates (slight generalizations
of the holomorphic anomaly equation \cite{Vafa}) fail.  The absence
of boundary contributions to the (chiral) $U(N)$ free energy
suggests an even simpler structure--that a handle creation
operator exists.  We find something akin to one, but it couples
to the infinitely many deformations of QCD.  A simple equation
in terms of the K\"{a}hler modulus alone would be much more
powerful. It is plausible,  but it has not been found.  So we
will list the exact expressions up to genus 8 and point out some
surprising features that emerge.

\section{QCD$_2$ and Its String Expansion}


This section presents an overview of some of the salient aspects of
QCD$_2$.  We give a brief review of the heat kernel partition function
and Gross's large $N$ expansion of it.  Next we discuss the
relationship between $SU(N)$ and $U(N)$ 2D Yang-Mills theory,
and we show that the $SU(N)$ partition function is easily
computed from that of chiral $U(N)$.

Pure glue Yang-Mills theory is exactly solvable in two dimensions.
This is largely due to the absence of propagating gluons,
as only global degrees of freedom survive gauge fixing.
There are no transverse gluons
to propagate.  The partition function is
\be
Z = \int [\DD A_\mu ] \, e^{-\frac{N}{4\lambda}
\int _\Sigma F_{\mu \nu} F^{\mu \nu} d^2x}
\label{QCDZF}
\ee
which may be calculated on any Riemann surface $\Sigma$.  It only
depends on the scaled gauge coupling $\lambda ~(=g_{QCD}^2 N)$,
the area $A$ of $\Sigma$,
the topology (genus $G$) of $\Sigma$ and the gauge group.  We will
consider Riemann surfaces with no boundaries.  The
gauge group will be either $SU(N)$ or $U(N)$, with a large $N$
in order to get the string expansion in $g_{st}=1/N$.

There is a remarkable solution of QCD$_2$
due to Migdal and Rusakov \cite{MR}.
The heat kernel lattice action reproduces (\ref{QCDZF}) in the
continuum limit, and it has the powerful feature that it is
renormalization group invariant.  This permits a quick solution
for the partition function
\be
Z = \sum _R (\dim R)^{2-2G} e^{-\frac{\lambda A}{2N} C_2(R)}
\label{QCDZ}
\ee
where $C_2(R)$ is the second Casimir and $\dim R$ is the dimension
of the representation $R$.  The sum over irreducible representations
may be expressed as a sum over different weights (i.e. Young tableaux).
A Young tableau has $n_k$ boxes in the
$k^{th}$ row and rows decreasing in length,
$n_1 \ge n_2 \ge \cdots \ge n_N$.
For $SU(N)$ each $n_k$ is a non-negative integer,
whereas for $U(N)$ the weights $n_k$ may be any integer.
The additional irreducible representations are due to the
$U(1)$ in $U(N) \cong (SU(N) \times U(1))/Z_N$.
The Casimir is
\be
\asize{2.0}
C_2(R)  = \left\{
\ball
{\dst
\sum _{k=1}^N n_k ( n_k + N + 1 - 2k)
- \frac{n^2}{N} } & {\rm ~~for~SU(N)} \\
{\dst \sum _{k=1}^N n_k ( n_k + N + 1 - 2k) }
& {\rm ~~for~U(N)}
\eac \right.
\asize{1.8}
\ee
where $n=\sum n_k$ and a particular choice is made for the
$U(1)$ charge in order to simplify the $U(N)$ Casimir.  (In
general, there is an extra term $\frac{\alpha n^2}{N}$, where
$\alpha$ is determined by the $U(1)$ charge
$\sqrt{\lambda ^{\prime}}$ in $U(N)$:
$\alpha = 1 - N \sqrt{\lambda^{\prime}/\lambda}$.)
The dimension of $R$ is given by
\be
\barcl
\dim R & = & {\dst
\prod _{1\le i < j \le N} \frac{n_i - n_j - i + j}{j - i} }.
\eac
\ee
It is a polynomial in $N$ of degree $n$,
\be
\barcl
\dim R & = & {\dst
\frac{1}{n!} \sum _{\sigma \in S_n} N^{K_\sigma} \chi _R(\sigma )
} \\ & = & {\dst
\frac{d_R}{n!} N^n \exp \left\{
\sum _{k=1}^\infty (-1)^k \frac{\Ctil _{(k+1)}(R)}{k\, N^k}
\right\} }
\eac
\label{dimse}
\ee
The first expression comes from
the Frobenius formula for characters $\chi _R(\sigma )$ of the
representation $R$ of the symmetric group $S_n$ \cite{GT}.
$K_\sigma$ is the number of cycles in the permutation $\sigma$.
The second expression may
be considered the large $N$ expansion of the standard
``factors over hooks'' rule for the dimension (cf. \cite{Georgi}).
The invariants $\Ctil _{(k)}(R)$ are given by
\be
\Ctil _{(k)}(R) = \sum _{j=1}^N \sum _{i=1} ^{n_j} (i-j)^{k-1}
\label{Ctil}
\ee
$\Ctil _{(k)}(R)$ is part of the $k^{th}$ Casimir, $C _{k}(R) =
k!\, \Ctil _{(k)}(R) + \cdots$.  For example,
$C _{2}^{U(N)}(R) = nN + 2 \Ctil _{(2)}(R)$ and
$n = \Ctil _{(1)}(R)$.
The resulting expression for the partition function is
\be
\barcl
Z & = & {\dst \sum _{n_i \ge n_{i+1}}
\prod _{1\le i < j \le N} \left(
{\tst \frac{n_i - n_j - i + j}{j - i}} \right) ^{2-2G}  \,
e^{-\frac{\lambda A}{2N} \sum _{k} \left[ n_k ( n_k + N + 1 - 2k)
- \frac{\alpha}{N} n^2 \right] } } \\
[2mm]
  & = & {\dst
e^{\frac{\lambda A}{24}(N^2-1)} \left\{ \sum _{h_i > h_{i+1} }
\prod _{1\le i < j \le N} \! \left( h_i - h_j\right) ^{2-2G}
e^{-\frac{\lambda A}{2N}[(\sum _{k} h_k^{\, 2}) -\frac{\alpha}{N}
n^2] } \right\}
\prod _{l=1}^{N-1} \left( l! \right) ^{2G-2} }
\eac
\ee
where $h_k = n_k + \half (N+1) - k$ (cf.\ \cite{DK}) and
$n=\sum n_k = \sum h_k$.

The $U(N)$ case is simpler both because of the form of the Casimir
and because the sums run over all integers $n_k$, including the
negative ones.  This lets us express the $G=1$ partition
function in terms of elliptic functions,
\be
\barcl
Z_{U(N)}^{G=1} & = &
{\dst \frac{1}{N!}
e^{\frac{\lambda A}{24}(N^2 - 1)}
\sum _{\stackrel{{\sst h_i \ne h_j}}{\forall i \ne j}}
e^{-\frac{\lambda A}{2N}\sum _k h_k^{\, 2} }} \\ [2mm]
& = & {\dst \frac{1}{N!} e^{\frac{\lambda A}{24}(N^2 - 1)}
\left\{ \vartheta (t) ^N
\asize{0.7}
- \left( \!
\bac N \\ 2 \eac
\! \right)
\asize{1.8}
\vartheta (t) ^{N-2} \vartheta (2t)
- \cdots
\right\} }
\eac
\label{thetaUN}
\ee
where $t= i\lambda A/2\pi N$.
The corrections come from terms in $\vartheta (t) ^N$ with $h_i = h_j$
for some $i \ne j$.  The Jacobi theta function $\vartheta$
is $\vartheta _2$ when $N$ is even and $\vartheta _3$ when $N$ is odd.

The $SU(N)$ partition function is more complicated.  Consider first
the $U(N)$ partition function sum restricted to $SU(N)$ Young
tableaux.  It may be expressed in terms of the functions
\be
\barcl
{\dst \vartheta _{(N)}(t) = \sum _{h=(3-N)/2}^{\infty}
e^{\pi i t h^2} } & = & {\dst
 \frac{1}{2} \vartheta (t)
+ \half \sum _{h=(3-N)/2}^{(N-3)/2} e^{\pi i t h^2} } \\
& = & {\dst \vartheta (t) - \sum _{h = (N-1)/2}^{\infty} e^{\pi i t h^2}}
\eac
\ee
Then the intermediate partition function is
\be
Z_{U/SU}  = \frac{1}{(N\! -\! 1)!}
e^{\frac{\lambda A}{24N}(N^3 - 3 N^2 + 5 N - 3)}
\left\{ \vartheta _{(N)} (t) ^{N-1}
\asize{0.7}
- \left( \!
\bac N \! - \! 1 \\ 2 \eac
\! \right)
\asize{1.8}
\vartheta _{(N)}(t) ^{N-3} \, \vartheta _{(N)}(2t)
- \cdots
\right\}
\label{ZUSU}
\ee
There is a related function, $Z^+$, in which the sum is restricted
to Young tableaux with fewer than $N/2$ boxes in any column.
\be
Z^+ = \frac{2^{-N/2}}{(N/2)!} e^{\frac{\lambda A}{48} (N^2 -1)}
\left\{ \vartheta _2 (t)^{N/2} - 2
\asize{0.7}
\left( \!
\bac N/2 \\ 2 \eac
\! \right)
\asize{1.8}
\vartheta _2 (t) ^{(N-4)/2} \, \vartheta _2 (2t)
- \cdots
\right\}
\label{Zchi}
\ee
(for $N$ even).
The large $N$ expansion of this seemingly ad hoc function will be
the focus of much of what follows.  Finally, to get the $SU(N)$
partition function, the extra term in the Casimir must be
included, but that is difficult.  In any case, these expressions
are curious, but they are not much help.  They work well for
small $N$:
\be
\barcl
{\dst Z_{U(1)}^{G=1} } & = &
{\dst \vartheta _{3} (t) } \\
{\dst Z_{U(2)}^{G=1} } & = &
{\dst \txthalf e^{\frac{\lambda A}{8}}
\left\{ \vartheta _{2} (t)^2
- \vartheta _{2} (2t)
\right\}  } \\
{\dst Z_{U(3)}^{G=1} } & = &
{\dst {\tst \frac{1}{6}} e^{\frac{\lambda A}{3}}
\left\{ \vartheta _{3} (t)^3
-3 \vartheta _{3} (t)
\vartheta _{3} (2t)
+2 \vartheta _{3} (3t)
\right\}  } \\
{\dst Z_{U(2)/SU}^{G=1} } & = &
{\dst {\tst \frac{1}{2}} e^{\frac{\lambda A}{6}}
\vartheta _{2} (t) } \\
{\dst Z_{U(3)/SU}^{G=1} } & = &
{\dst {\tst \frac{1}{2}} e^{\frac{\lambda A}{6}}
\left\{ \vartheta _{3} (t)^2
+2 \vartheta _{3} (t)
-2 \vartheta _{3} (2t) -1
\right\}  } \\
{\dst Z_{SU(2)}^{G=1} } & = &
{\dst {\tst \frac{1}{2}} e^{\frac{\lambda A}{8}}
\left\{ \vartheta _{3} (t/2) -1 \right\}  }
\eac
\label{Nsmall}
\ee
Unfortunately, (\ref{thetaUN}) and (\ref{Zchi})
are not conducive to large $N$ expansions, since it is difficult
to determine if one term dominates the sum.  Also, note that even
the simpler $U(N)$ partition function expressed in
terms of elliptic functions is not a modular form,
since each term has a different weight.  We will see below
that at each worldsheet genus the $G=1$ string free energy
is almost a modular form, but with a different modulus
$\tau = \frac{N}{2} t$.
On the other hand, the modular weights of the
theta functions do determine the small area behavior of the
partition function.
Since $\vartheta _3 (t) \sim t^{-\half}$ as $t \ra 0$, the leading
term dominates the small area limit at finite $N$.  If $N$ is then
taken to infinity, the partition function develops an essential
singularity at $t=0$.  It is not clear from (\ref{thetaUN}) if one
term dominates in the large $N$ limit, which should be taken first.
We will see below that the string partition
function has an essential singularity at $\lambda A=0$ which is
the phase transition that occurs at finite coupling on the
sphere \cite{DK}.


The string expansion for $Z$ is a large $N$ expansion.  This is
explained in detail in \cite{Gross} and \cite{GT}, but a brief
discussion of the structure will help motivate the ensuing analysis.
Gross and Taylor have shown how $SU(N)$ representations with
relatively small Casimirs (of order $N$) and small dimensions
give the leading
contribution to the partition function at large $N$, yielding
a series that has many properties of closed string
perturbation theory.  A Young tableau with a small number of
boxes has a relatively small Casimir, and it makes a leading
contribution.  But for $SU(N)$, a representation $R$ and its
complex conjugate $\Rbar$ have the same Casimir and the same
dimension, so they make the same contribution to $Z$.  This
leads to a natural factorization of representations.  The
``chiral'' representations are those with no more than $N/2$
boxes in any column, and the ``anti-chiral'' representations are
those whose complex conjugate is chiral with no column of
$N/2$ boxes.\footnote{This is a
slight generalization of the definition given by Gross and
Taylor \cite{GT}.}  (Recall that if $R$ has $c_j$ boxes in
its $j^{th}$ column, then $\Rbar$ has $N-c_j$ boxes in its
$j^{th}$ column from the right.)  Then any representation is
expressed uniquely as the Young product of an anti-chiral and
a chiral representation; i.e.\ any tableau is a chiral
tableau joined to an anti-chiral tableau.

The physical partition function may be obtained from the chiral
$U(N)$ partition function, $Z^+$, in which the sum is restricted to
chiral $SU(N)$ representations,
\be
Z^+ = \sum _{R^+} (\dim R) ^{2-2G} e^{-\frac{\lambda A}{2N}
2 \Ctil _{(2)} } e^{-n \Atil}
\ee
This is a well-defined function whose asymptotic expansion is
the holomorphic topological string perturbation theory.
$\Atil= \half \lambda A$, but it
is kept formally independent of $A$ so that the extra
piece of the $SU(N)$ Casimir may be obtained by differentiation
with respect to $\Atil$.
For example, the chiral $SU(N)$ partition function is given by
\be
Z^+ _{SU(N)}
= \left. e^{(\ANN ) \d _{\! \Atil}^{\, 2}} \, Z^+ (\Atil )
\right| _{\Atil  = \half \lambda A} .
\label{SUfromU}
\ee
The free energy $F=\log Z$ is
\be
\barcl
F^+ _{SU(N)} & = & {\dst F^+
+ \sum _{m=1}^{\infty} \Ptil _m \left(x_j =
P_{2j} (\fp ,\fpp ,\cdots , F^{+(2j)})/j! \right) (\ANN )^m } \\
& = & {\dst
F^+
+ {\tst (\ANN )} \left[ \fpp + (\fp )^2 \right]
} \\ & & ~~~~~ {\dst
+ \frac{(\ANN )^2}{2!}\left[ \fpb +4 \fp \fpbb + 3 (\fpp )^2
+ 6 (\fp )^2 \fpp + (\fp )^4 \right]  + \cdots }
\eac
\label{fSUfromU}
\ee
where the Schur polynomials $P_n$ are generated by
$e^{x_kz^k} = \sum P_n(x_1,\cdots ,x_n) z^n$, the
polynomials $\Ptil _n$ are generated by
$\log (1+x_kz^k) = \sum \Ptil _n(x_1,\cdots ,x_n) z^n$,
and the primes denote $\d _{\Atil}$.
The full $SU(N)$ partition function on the torus is
\be
Z_{SU(N)}^{(G=1)}
= \left. e^{(\ANN ) (\d _{\! \Atil _1}- \d _{\! \Atil _2})^{\, 2}}
\, Z^+(\Atil _1) \, Z^+(\Atil _2)
\right| _{\Atil _1 = \Atil _2 = \half \lambda A}
\label{fullSU}
\ee
using the formula for the Casimir of the Young product of a chiral
and an anti-chiral representation, $C_2(\Sbar R) = C_2(R) + C_2(S)
+ 2n\ntil /N$.  The $G=1$ free energies are related by
\be
\barcl
F & = & {\dst 2 F^+
+ \sum _{m=1}^{\infty} \Ptil _m \left(x_j =
{\tst \frac{1}{j!}}P_{2j} (0 ,2\fpp ,0 ,2F^{+(4)},\cdots , 2F^{+(2j)})
\right) (\ANN )^m } \\
& = & {\dst
2 \left\{ \rule{0in}{.25in} F^+ + {\tst (\ANN )} \fpp
\right. } \\ & & ~~~~~ {\dst \left.
+ \frac{(\ANN )^2}{2!}\left[ \fpb + 6 (\fpp )^2 \right]  +
\frac{(\ANN )^3}{3!} \left[ \fpc + 30 \, \fpp \, \fpb + 60 (\fpp )^3\right]
\right. } \\ & & ~~~~~ {\dst \left.
+ \frac{(\ANN )^4}{4!} \left[ \fpd + 56 \, \fpp \, \fpc + 70 (\fpb )^2 +
840 \left( (\fpp )^4 + (\fpp )^2 \fpb ) \right) \right] + \cdots \right\} }
\eac
\label{fullfSU}
\ee
Note that only even derivatives of $F^+$ enter (\ref{fullfSU}).
As a result $F^+$ and $F$ (but not $F^+_{SU(N)}$ or $F_{U(N)}$)
have a simple modular structure independent of the zero point
energy.
Once the chiral $U(N)$ free energy is known, it is straight-forward
to calculate $F^+ _{SU(N)}$ and $F _{SU(N)}$.  Of course, it is
trivial to get $F^+ _{U(N)}$ from the full $SU(N)$ free energy,
$F _{SU(N)}$, because of its modular structure.

The form of (\ref{fullSU}) suggests that for $G=1$ the $U(N)$
chiral partition function is on the same footing as the more
physical non-chiral $SU(N)$ partition function.  This is remarkable
since a topological string theory reproducing simple Hurwitz space
(i.e.\ the moduli space of chiral $U(N)$, cf. \cite{GMtop}) is
relatively easy to construct.  There are no contributions from the
boundary of moduli space;  the contact terms vanish.  A very
simple, explicit perturbation gives the full theory.  This is in
striking contrast to the complications of $G\ne 1$.

The formula for the dimension of a composite
representation is not especially simple, so
(\ref{fullSU}) and (\ref{fullfSU}) require unwieldy
corrections for $G\ne 1$.
In fact, of all the group invariants, only the quadratic
Casimir has such a simple decomposition.  The higher Casimirs,
$C_k(\Sbar R)$, decompose into a sum of products of the lower
Casimirs $C_l(R)$ and $C_l(S)$ with $l\le k$.
The full partition function of QCD
perturbed by $C_k$, may be expressed as a deformation of two
copies of $Z^+$ depending on the couplings of all the lower
$C_j$'s.  The higher the Casimir, the more couplings that must be
differentiated.  The dimension of $R$ may be expressed in terms of
the $C_j$'s as well, but in the large $N$ limit it takes infinitely
many, so the analog of (\ref{fullSU}) for $G\ne 1$ requires
derivatives with respect to infinitely many couplings.

The perturbed chiral $U(N)$ partition function may be written
\be
Z^+(\Atil , \Atil _2, \cdots ) = \sum _{R^+} (\dim R)^{2-2G}
\, \exp \left( \sum _k \Atil _k C_{k}(R) /N^{k-1}
\right) e^{-n\Atil} .
\ee
This partition function results from the
chiral reduction of a renormalization group invariant heat
kernel lattice action as before, but it leads to perturbations
of the Yang-Mills action $F^2$ by higher powers of the field
strength $(F^2)^k$ in the continuum.  Its large $N$
expansion is string-like, since only even powers of $1/N$
arise.  This results from a cancellation in the sum over
representations.  It follows from the definition (\ref{Ctil})
that $\Ctil _{(k)}(R) = (-1)^{k-1} \Ctil _{(k)}(S)$,
if the columns of $R$'s Young tableau are the rows of $S$'s.
Since the Casimir $C_{k}(R)$ is a homogeneous polynomial in
$\Ctil _{(l)}(R) /N^{l-1}$, the relative minus sign cancels any
occurrence of an odd power of $1/N$ arising from the perturbations
or the dimension (\ref{dimse}).  This argument immediately
extends to the $SU(N)$ case and the full partition function.
Of course, there are other properties a string perturbation
expansion should possess, but these will be verified elsewhere
\cite{QCDlo}.


The large $N$ chiral partition function for $U(N)$ Yang-Mills theory
on the torus has been shown to be the following sum over classes
of $n$-sheeted holomorphic covering maps $\nu _n$
\cite{GT}
\be
\barcl
F^+ & = & {\dst
\sum _{g=G}^{\infty} N^{2-2g}
\sum _{n=1}^{\infty} ~ \sum _{[\nu _n]: \Sigma _g \ra \Sigma _{G=1}}
f_{g,\nu _n}^G \, (\txthalf \lambda A)^i \, e^{-\half \lambda nA} }
\eac
\label{exQCDpar}
\ee
where $\Sigma _g$ ($\Sigma _G$) is the genus g (G) worldsheet
(space-time) Riemann surface, and $[\nu _n]$ is a class of
$n$-sheeted covers with $i$ square root
branch points.
$f_{g,\nu _n}^G$ is the number of holomorphic maps divided by
a symmetry factor characterizing the moduli space of maps.
The maps satisfy the Riemann-Hurwitz relation
\be
2(g\! -\! 1) = 2n(G-1) +i .
\label{RieHur}
\ee
Since $i=2g-2$ for maps to the torus, the factors multiplying
the exponential in (\ref{exQCDpar}) combine into
$(\lambda A/2N)^{2g-2}$.
The QCD partition function on other ($G\ne 1$) Riemann surfaces
is more complicated than (\ref{exQCDpar}).
Even the $SU(N)$ partition function on the torus is more complicated,
with contributions coming from the boundary of moduli space
where worldsheet handles are collapsed to a point in space-time
\cite{Min}.
For $G\ne 1$ at $\lambda A=0$, the free energy is a sum of the orbifold
Euler characteristics of the moduli spaces \cite{GMtop}.



We will study QCD on the torus, so there is no contribution to
the free energy coming from genus 0 worldsheets.  The lowest
genus worldsheet to contribute is an unbranched single cover of the
target space, so $g_{min} = G$ according to the Riemann-Hurwitz
formula.  Since $F_0 = 0$ the relationship between the free
energy and the partition function is simple:
\be
\barcl
Z & = & {\dst
e^{F_1} \exp \left\{ \sum _{g=2}^{\infty} (1/N)^{2g-2} F_g \right\} } \\
  & = & {\dst
e^{F_1} \left[ (1/N)^2 F_2 + (1/N)^4 \left(F_3 + {\txthalf}
F_2^{\, 2} \right)
+ \cdots \right] .}
\eac
\label{ZtoF}
\ee
We will usually discuss the free energy below, but the partition
function is very similar in form.  For instance, when $F^+_g$
has modular weight $6g\! -\! 6$, so does $Z^+_g$ apart from the
factor $e^{F^+_1} = \eta$.  This eta function factor gives the
partition function an essential singularity at zero area/coupling
which can cause difficulties (like order of limits problems)
that do not arise with the free energy.

\section{QCD On The Torus}

In this section we will proceed to study some properties of
the $G=1$ free energy in detail.  The free energy on the torus has
an interesting structure reminiscent of critical string theory,
but not shared with QCD on other Riemann surfaces.  It transforms
nicely under target space modular transformations, as expected
for a string theory.  It is not exactly modular covariant, but it
turns out that $F^+_g$ is an anomalous modular form of weight
$6g\! -\! 6$.  The deviations from modular covariance are of an
interesting form, and they conspire to give an unexpectedly mild
behavior at small area
and/or weak gauge coupling.  This may be related to a hidden
symmetry in the string theory.  The section begins with the
derivation of a generating function for $F^+_g$, followed by
a general description of its modular properties.  Then both
are used to calculate $F^+_g$ and $F_g$ for $g=1,\cdots ,8$,
exactly in terms of modular forms.  Finally, we examine the
large and small area behavior of the free energy.

QCD on the torus is simpler in many ways than on the sphere or
on higher genus surfaces.  The dimension factor
drops out of the heat kernel expression for the partition function
(\ref{QCDZ}):
\be
Z^+ = \sum _{n_1 \ge n_2 \ge \ldots \ge n_{N/2} \ge 0}
e^{\AN \sum _{k} [n_k (n_k + 1 -2k)]}
\, e^{-n \Atil} .
\label{Zgone}
\ee
In the Gross-Taylor description this means that no
omega points or omega-inverse points are required.  The chiral
partition function just counts simply branched maps.  Despite the
absence of omega points, the free energy is non-trivial since
the map $\nu : \Sigma _g \ra \Sigma _{G=1}$ can wrap arbitrarily
many times.  The wrapping number $n$ drops out of the
Riemann-Hurwitz formula (\ref{RieHur}) for $G=1$.  This allows
non-trivial modular transformations of the K\"{a}hler modulus
$\Atil$.  In addition to formulations of QCD$_2$ using gauge theory
and string theory techniques that work for any Riemann surface,
QCD on the torus
has been reformulated as a two dimensional free fermion theory and a
Jevicki-Sakita boson theory \cite{Douglas,MP}.  The relative
simplicity of QCD on the torus along with the alternative formulations
allows us to say a great deal about the string theory.

The partition function (\ref{Zgone}) is similar to a theta
function, as we saw above in (\ref{thetaUN}).  It would be nice
to extract the large $N$ expansion directly from the theta functions,
but that has proved to be difficult.
We will now develop a way to sum the series
(\ref{Zgone}) that is more amenable to a large $N$ expansion.
Since the sum is restricted to chiral representations, the
factor of $1/N$ appearing in the first
exponential in (\ref{Zgone}) is the string coupling.
$Z^+$ may be expressed in terms of a generating function:
\be
\barcl
Z^+ & = & {\dst \left. \sum _{n_1 \ge n_2 \ge \ldots \ge n_{N/2} \ge 0}
e^{\AN \sum _{k} [(-\d _{\beta _k}) (-\d _{\beta _k} + 1 -2k)]}
\, e^{-\sum _{k} \beta _k n_k} \right| _{\beta _j = \Atil}
} \\
& = & {\dst \left. \sum _{m_j \ge 0 ,~ \forall j}
e^{\AN \sum _{k}
\left( k \d _{\alpha _k}^{\, 2} + k^2 \d _{\alpha _k}
+ 2k \d _{\alpha _k} \sum _{l=k+1}^{\infty} \d _{\alpha _l} \right) }
\, e^{-\sum \alpha _k m_k}
\, \right| _{\alpha _j = j \Atil}  }
\eac
\ee
where $m_k = n_k - n_{k+1}$
(i.e.\ $n_k = \sum _{l=k}^{\infty} m_l$),
$\beta _k = \alpha _l - \alpha _{l-1}$ and
$\d _{\beta _k} = \sum _{l=k}^{\infty} \d _{\alpha _l}$.
All of the sums over the $m_j$ begin at zero, so the generating
function is just a product of geometric sums.  Summing these
series (and allowing for a zero point energy $\epsilon _0$)
we find
\be
\barcl
{\dst Z^+} & = & {\dst
\left. e^{\epsilon _0 \Atil/24} \,
\exp \left\{ \AN \sum _{k=1}^{\infty}
\left( k \d _{\alpha _k}^{\, 2} + k^2 \d _{\alpha _k}
+ 2k \d _{\alpha _k} \sum _{l=k+1}^{\infty} \d _{\alpha _l} \right)
\right\} \, \frac{1}{\etatil (\alpha )}
\, \right| _{\alpha _j = j \Atil}
} \\ & = & {\dst
\left. e^{\epsilon _0 \Atil/24} \,
\exp \left\{ \AN \sum _{k=1}^{\infty}
\left( k^2 \d _{\alpha _k} + \d _{\Atil} \d _{\alpha _k}
- \half \sum _{l=1}^{\infty} \left| k-l \right|
\d _{\alpha _k} \d _{\alpha _l} \right)
\right\} \, \frac{1}{\etatil (\alpha )}
\, \right| _{\alpha _j = j \Atil}
}
\eac
\label{etadiff}
\ee
where the generalized eta function is given by
\be
\frac{1}{\etatil (\alpha )} = e^{\Atil /24}
\prod _{j=1}^{\infty} \frac{1}{1-e^{-\alpha _j}} + \cdots .
\label{eta}
\ee
The omitted terms correct for the fact that we have allowed
an infinite number of rows instead of stopping $j$ at $N/2$, the
maximum for a chiral tableau.  These terms are proportional
to some power of $e^{-N}$, so they are exponentially small
non-perturbative corrections.  We can ignore them.
When (\ref{eta}) is
evaluated at $\alpha _j = j \Atil $ in the large $N$ limit, the
generalized eta function becomes the usual eta function
\be
\barcl
{\dst \left. \etatil \, \right| _{\alpha _j = j \Atil} }
& {\dst \map{N \ra \infty} } &
{\dst \eta } ~~~ {\dst = q^{1/24} \prod _{n=1}^{\infty} (1-q^n) }
\eac
\ee
with $q = e^{-\Atil}$ (so that the K\"{a}hler modulus of the torus
is $\tau = -\Atil / (2 \pi i)$).
This is the genus $g=1$ chiral partition function for
$SU(N)$ and $U(N)$ Yang-Mills on the torus found in \cite{Gross}.

The free energy is the ``connected'' part of (\ref{etadiff}).
It may be computed using the Campbell-Baker-Hausdorff formula.
This is particularly simple because $\log \eta (\alpha) =
-\frac{\Atil}{24} + \sum _j \log (1-e^{\alpha _j})$, and
different derivatives cannot act on the same term in the sum.

It is key for a closed string formulation of QCD$_2$ that
the odd powers
of $1/N$ vanish in the partition function.  This leads to some
non-trivial identities\cite{Gp}.  For example, the term in
(\ref{etadiff}) proportional to $1/N$ (corresponding to
genus $g=3/2$) is
\be
\sum _{k=1}^{\infty} \left\{ -\frac{k(k-1)\, q^k}{1-q^k}
+ \frac{2k\, q^{2k}}{1-q^{2k}}
+ \sum _{l=1}^{\infty} \frac{(k+l)\, q^{k+kl+l}}{(1-q^k)(1-q^l)}
\right\} = 0 .
\label{oddgid}
\ee
The argument given above for the absence of these terms,
based on Gross's original discussion, is essentially group
theoretic.  It would be interesting to
understand the identity (\ref{oddgid}) and its generalizations
from the point of view of number theory.

The free energy is almost, but not quite, invariant under modular
transformations.  Already we have seen that $F_1^+ = - \log \eta$.
This is not modular invariant, contrary to what one would expect
for a string theory.  Under the transformation of the K\"{a}hler
modulus $\Atil \ra 4\pi ^2 / \Atil$, the free energy has a
modular anomaly:
$\log \eta (4 \pi ^2 / \Atil ) =
\log \eta (\Atil ) + \txthalf \log (\Atil /2 \pi )$.
It becomes modular invariant if we make an ansatz that the theory
has a holomorphic anomaly.\footnote{The holomorphic anomaly arises
in topological string theory, coming from the contribution of
BRST exact operators (with anti-holomorphic couplings)
at the boundary of moduli space.
\cite{Vafa}}  The anomaly makes a contribution to $F^+$ that
depends on $\bar{\Atil}$ as well as $\Atil$.  Then
\be
F_1^+ = -\log \left\{ \sqrt{\txthalf (\Atil + \bar{\Atil})} \,
\eta (\Atil ) \, \eta (\bar{\Atil}) \right\} .
\ee
This is modular invariant, and the original
free energy is recovered by sending $\bar{\Atil} \ra \infty$,
up to an infinite shift in the zero point energy.
Modular invariance is an extremely useful property.  It is a kind
of strong/weak coupling duality in the gauge coupling $\lambda$,
and it determines
the general structure of the free energy in terms of modular forms.

The higher genus free energies are also expected to be modular
covariant after the non-holomorphic completion.  The string coupling
transforms non-trivially (with weight $-1$), so the higher genus
free energies should transform with a definite modular weight.
Douglas \cite{Douglas} was able to show that this is the
case using his formulation of QCD$_2$ on the torus in
terms of a Jevicki-Sakita two dimensional boson system:
\be
Z^+ = \int \DD \varphi \, e^{-\int \d \varphi \dbar \varphi +
g_{st} A (\d \varphi)^3} .
\ee
The propagator is $\left< \d \varphi (z) \, \d \varphi (0) \right>
= - \d ^2 \log \vartheta _1 (z|\Atil )
 - \frac{4\pi ^2}{\Atil + \bar{\Atil}}
= \wp (z|\Atil ) + \pi ^2 E_2 (\Atil )/ 3
 - \frac{4\pi ^2}{\Atil + \bar{\Atil}}$,
where $\vartheta _1$ is the Jacobi theta function and the
Weierstrass $\wp$ function is given by
$\wp (z|\Atil ) = z^{-2} + \sum _k 2 (2k-1) \, \zeta (2k) \,
E_{2k}(\Atil ) \, z^{2k-2}$.
The functions $E_{2k}$ are the Eisenstein series--weight
$2k$ modular forms.  They are described below.
The point is that the propagator has modular weight two,
where a weight $k$ modular form
transforms as $M_k (-1/\tau ) = \tau ^k M_k (\tau )$.
This determines the weight of the genus
$g$ free energy
\be
F_g^+ = \frac{1}{(2g\! - \! 2)!}
\left( \frac{\lambda A}{2N}\right) ^{2g-2}
\left< \left[ \int :(\d \varphi)^3: \right] ^{2g-2} \right> _{connected}
 .
\ee
It has $3g\! -\! 3$ propagators, giving a total weight $6g\! - \! 6$.

It would be fantastic if we could do the integrals to determine
which weight $6g\! - \! 6$ modular form $F_g^+$ is.
Unfortunately, these integrals are very difficult.  Douglas has
done the $F^+_2$ integral \cite{Douglas4}, and we have done the
$F^+_3$ integrals.  The higher genus integrals are extremely
difficult, since the convolution of offset $\wp$ functions is
not elliptic.  Also, as the genus $g$ increases, there are
a growing number of diagrams and integrals.  In fact,
the number of diagrams at genus $g$ is proportional to $g!$, as
easily seen in the zero dimensional $\varphi ^3$ integral.
In addition, ever higher weight modular forms entering from the
$\wp $ function must be reduced.
It does not seem promising to calculate the higher genus free
energies this way.  Fortunately, the fact that they are modular
forms determines them up to a few coefficients.
These coefficients may be found by
taking parametric derivatives of the generalized
$\eta $ function (\ref{etadiff}).

A set of $3g^2/4$ ($(3g^2+2g-5)/4$ for odd $g$) modular forms
comprises a basis at weight $6g\! - \! 6$.  The basis is
made of products of $E_2, E_4$ and $E_6$ or
$E_2, \ep$ and $\epp$, which have weight 2, 4 and 6, respectively.
Most texts on modular forms write the basis in terms of
$E_4$ and $E_6$, but for our purposes the second
basis is preferable.
The two bases are interchangeable, since
$E_4 = (E_2)^2 + 12 E_2^{\, \prime}$ and
$E_6 =  E_2^{\, 3} + 18 E_2 \, E_2^{\, \prime} +
36 E_2^{\, \prime \prime}$.  Also, note that
$E_2^{\, \prime \prime \prime} =
\frac{3}{2} [\ep ]^2 - \e \epp$ and
$E_{k+2} = E_k E_2 + \frac{12}{k} E_k^{\, \prime}$ for
$k=4,6,\cdots$.  The prime means differentiation
with respect to $\Atil$, so $f^{\prime} = \d _{\Atil} f =
\frac{-1}{2\pi i}\d _{\tau} f$, and for finite $\bar{\Atil}$
it becomes the covariant derivative
\be
D_{\Atil} = \d _{\Atil}
+ \frac{k}{\Atil + \bar{\Atil}}.
\ee
acting on a weight $k$ modular form.
Regardless of which basis
is used, the genus $g$ free energy is determined up to roughly
$3g^2/4$ coefficients
\be
F_g^+ = \left( \frac{\lambda A}{2N}\right) ^{2g-2~~}
\sum _{k=0}^{3g-3} \sum _{2l+3m=3g-3-k} c_{kl} \, E_2^{\, k} \,
(E_2^{\, \prime})^l \, (E_2^{\, \prime \prime})^m .
\label{FEs}
\ee
The coefficients $c_{kl}$ are rational numbers to be
determined.

The $k^{th}$ Eisenstein series, $E_k$, is
\be
\barcl
E_k & = & {\dst
\frac{1}{2 \zeta (k)} \left. \sum _{m,n} \right. ^{\prime}
\frac{1}{(m\tau + n)^k} } \\
& = & {\dst
1 - \frac{2k}{B_k} \sum _{n=1}^{\infty} \frac{n^{k-1}\, q^n}{1-q^n}
= 1 - \frac{2k}{B_k} \sum _{n=1}^{\infty} \sigma _{k-1}(n)\, q^n
}
\eac
\label{Eser}
\ee
for $k \in 2{\bf Z}^+$ (cf.\ \cite{Kob}).  The number theoretic function
$\sigma _{k}(n)$ is the sum of the $k^{th}$ power of the
divisors of $n$, $\sigma _{k}(n)= \sum _{d|n} d^k$, and $B_k$ is the
$k^{th}$ Bernoulli number.  Every $E_k$ is a modular form of weight $k$,
except $E_2$.  There is no modular form of weight two,
although $E_2$ comes close.  The Eisenstein series transform as
\be
E_2(-1/\tau ) = \tau ^2 \left( E_2(\tau ) + \frac{12}{2 \pi i \tau}
\right)
 ~~~~~~E_k(-1/\tau ) = \tau ^k E_k(\tau ) ~~~k=4,6,\cdots
\ee
The fact that the higher Eisenstein series are modular forms is
easily seen from the definition (\ref{Eser}) since the
sums converge absolutely (and uniformly) on the upper half plane.
Since $E_2$ has a modular anomaly, it must be covariantized by
adding a term depending on $\bar{\Atil}$
\be
E_2 = 24 \, \d _{\Atil} F_1^+ = 1 -24 \sum _{n=1}^{\infty}
\sigma _{1}(n)\, q^n - \frac{12}{\Atil + \bar{\Atil}}
\ee
so $E_2 = 1 - 24 e^{-\Atil} - 72 e^{-2\Atil} - \cdots$.

The $G=1$ free energy for chiral $U(N)$ (restricted to $SU(N)$ tableaux)
has been calculated exactly up to worldsheet genus 8.  It would
be possible to continue up to genus 11, but beyond that the
computation takes too long even on fast computers.  A glance
at the genus 8 result (\ref{feight}) reveals that these
expressions would fill pages.
The first few free energies are given by
\be
F_1^+ = \frac{\epsilon _0}{24} \Atil - \log \eta
\ee
\be
\barcl
F_2^+ & = & {\dst
\frac{\gst ^2}{2!\, 2^4\, 3^4\, 5}\left\{ 5 E_2^{\, 3}
-3 E_2 E_4 -2 E_6 \right\} } \\
& = & {\dst - \frac{\gst ^2}{2!\, 90}\left\{\e \ep + \epp \right\}}
\eac
\ee
and
\be
\barcl
F_3^+ & = & {\dst \frac{\gst ^4}{4!\, 2^7\, 3^6} \left\{ -6 E_2^{\, 6}
     +15 E_2^{\, 4} E_4
     +4 E_2^{\, 3} E_6 -12 E_2^{\, 2}E_4^{\, 2} -12 E_2 E_4 E_6
     +7 E_4^{\, 3} +4 E_6^{\, 2} \right\} } \\
& = & {\dst \frac{\gst ^4}{4!\, 54} \left\{ 7 \epb ^3 +3 \eppb ^2 \right\}}
\eac
\ee
$F^+_1$ and $F^+_2$ have been calculated previously by Gross \cite{Gross}
and by Douglas \cite{Douglas4}, respectively.  The higher free energies
were not known.  The remaining expressions $F^+_4, \cdots , F^+_8$ are
increasingly lengthy, so they are listed in the appendix.

Using (\ref{fSUfromU}) and (\ref{fullfSU}) these results for the chiral
$U(N)$ free energies may be converted into the more complicated chiral
$SU(N)$ and full $SU(N)$ free energies.  The additional terms in the
full $SU(N)$ free energy coming from turning off the $U(1)$ coupling
and combining the chiral sectors are still modular forms
(up to the $E_2$ modular anomaly), but they have lower weights.
The corrections for the chiral $SU(N)$ free energy do not have
a definite weight because $\fp = (E_2 -1)/24$ enters (\ref{fSUfromU}).
The same is true of the non-chiral $U(N)$ free energy.
The general form of the full $SU(N)$ free energy (the
analog of (\ref{FEs}) for chiral $U(N)$) is
\be
F_g = g_{st}^{~2g-2} \sum _{l=0}^{g-1} (\lambda A)^{2g-2-l}
\sum _{k=0}^{3g-3-l} E_2^{\, k} \, M_{6g-6-2l-2k,l}
\ee
where $M_{k,l}$ is a true modular form of weight $k$ or a
weight $k$ combination of $\ep$ and $\epp$ in the other basis.
For example,
\be
\barcl
{\dst F_3} & = &
{\dst \frac{\gst ^4}{4!\, 27} \left\{ 7 \epb ^3 +3\eppb ^2\right\}
- \frac{\gst ^2 \gsu }{4!\, 3} 4 \ep \epp
} \\ & & ~~~~~~~~~~~~~{\dst
+ \frac{\gsu ^2}{4!\, 4} \left\{
4 \e \epp - 5 [\ep ]^2 \right\} }
\eac
\ee
$F_1$ through $F_4$ are given in the appendix.  The higher $SU(N)$
free energies are omitted for brevity.

A surprise about the free energy is that it has an unexpectedly
mild singularity as $\lambda A \ra 0$.  To see this
it will be useful to know the large and small area
behavior of the Eisenstein series.  The series are normalized
so that
\be
E_k \ra 1 + \OO (e^{-\Atil}) ~~~~{\rm as}~~\Atil \ra +\infty
\ee
for $k=2,4,\cdots $.
The small area limit is related to this by a modular transformation
\be
\barcl
E_k & \ra & {\dst \tau ^{-k} + \OO (e^{-2 \pi i / \tau })
 ~~~~{\rm as}~~\tau \ra i 0^+ ~~~(k=4,6,\cdots) } \\
 & \ra & {\dst (-\Atil /2\pi i) ^{-k} + \OO (e^{-4 \pi ^2 / \Atil })
 ~~~~{\rm as}~~\Atil \ra  0^+ } \\
E_2 & \ra & {\dst \frac{1}{\tau ^{2}} - \frac{12}{2 \pi i \tau}
+ \OO (e^{-2 \pi i / \tau }) ~~~~{\rm as}~~\tau \ra i 0^+} \\
 & \ra & {\dst \frac{-4 \pi ^2}{\Atil ^{2}}
 + \frac{12}{\Atil}  + \OO (e^{-4 \pi ^2 / \Atil })
 ~~~~{\rm as}~~\Atil \ra  0^+ }
\eac
\ee
It is amusing to note that these formulas also result from using
the Euler-Maclaurin formula on the $q$-expansion (\ref{Eser}).
The usual derivation of the modular transformation laws is
quite different.

Consider the expression for the partition function (\ref{etadiff}).
In the large area limit $F_g^+$ goes like $e^{-2\Atil }$ for $g\ge 2$.
There is no constant term, which would come from non-covering maps.
Similarly there is no single cover term proportional to $e^{-\Atil}$,
since higher genus surfaces must be at least double covers of the
torus.  The absence of the constant term shows up in the expressions
for $F^+_g$ since
each term contains a factor of $\ep$ or $\epp$ which begin with
$e^{-\Atil}$.  It is part of the
reason that the expressions are simpler in this basis.  The small
$\Atil$ limit is much more interesting.  A weight $6g-6$ modular
form would go like $(1/\Atil ) ^{6g-6}$ as $\Atil \ra 0^+$.
A much softer singularity is observed.  Up to terms of order
$\OO (e^{-1/\Atil })$ the following expressions hold:

\asize{1.7}


\be
F_2^+ =
\left( \AN \right) ^2 \left\{
\frac{2}{3\, \Atil ^3}
- \frac{2\, {\pi ^2}}{3\, {\Atil ^4}}
+ \frac{8\, {\pi ^4}}{45\, {\Atil ^5}}
\right\}
\ee

\be
F_3^+ =  \left( \AN \right) ^4 \left\{  \frac{-8}{{\Atil ^6}}
+ \frac{16\, {\pi ^2}}{{\Atil ^7}}
- \frac{100\, {\pi ^4}}{9\, {\Atil ^8}}
+ \frac{224\, {\pi ^6}}{81\, {\Atil ^9}}
\right\}
\ee

\be
F_4^+ = \left( \AN \right) ^6 \left\{    \frac{2272}{9\, {\Atil ^9}}
- \frac{2272\, {\pi ^2}}{3\, {\Atil ^{10}}}
+ \frac{8096\, {\pi ^4}}{9\, {\Atil ^{11}}}
- \frac{41504\, {\pi ^6}}{81\, {\Atil ^{12}}}
+ \frac{48256\, {\pi ^8}}{405\, {\Atil ^{13}}}
\right\}
\ee

\be
\barcl
F_5^+ & = & {\dst \left( \AN \right) ^8 \left\{   \frac{-13504}{{\Atil ^{12}}}
+ \frac{54016\, {\pi ^2}}{{\Atil ^{13}}}
- \frac{834304\, {\pi ^4}}{9\, {\Atil ^{14}}}
+ \frac{7010816\, {\pi ^6}}{81\, {\Atil ^{15}}}
\right. } \\ & & {\dst ~~~~~~~~~~~~~~~~~ \left.
- \frac{17887904\, {\pi ^8}}{405\, {\Atil ^{16}}}
+ \frac{11958784\, {\pi ^{10}}}{1215\, {\Atil ^{17}}}
\right\} }
\eac
\ee

\be
\barcl
F_6^+ & = & {\dst \left( \AN \right) ^{10}
\left\{\frac{15465472}{15\, {\Atil ^{15}}}
- \frac{15465472\, {\pi ^2}}{3\, {\Atil ^{16}}}
+ \frac{105156608\, {\pi ^4}}{9\, {\Atil ^{17}}}
- \frac{418657280\, {\pi ^6}}{27\, {\Atil ^{18}}}
\right. } \\ & & {\dst ~~~~~~~~~~~~~~~~~~ \left.
+ \frac{572409344\, {\pi ^8}}{45\, {\Atil ^{19}}}
- \frac{2467804672\, {\pi ^{10}}}{405\, {\Atil ^{20}}}
+ \frac{33778284544\, {\pi ^{12}}}{25515\, {\Atil ^{21}}}
\right\} }
\eac
\ee
\asize{1.8}
The expressions for $F_7^+$ and $F_8^+$ are easily calculated
as well from (\ref{fseven}) and (\ref{feight}).

In general the small area behavior of the free energy is
\be
F_g^+ =  \left( \AN \right) ^{2g-2} \sum _{k=3g-3}^{4g-3}
\frac{c_{k,g} \pi ^{2(k-3g+3)}}{\Atil ^k} +
\OO (e^{-1/\Atil })
\ee
where each $c_{k,g}$ is a rational number.
This form may be proven using the Campbell-Baker-Hausdorff
formula to get $F_g^+$ from $Z^+$ (\ref{etadiff}), and then
using the Euler-Maclaurin formula to replace the sums with
integrals.  This small area
behavior is interesting for a number of reasons.  It might be
the result of a symmetry in QCD$_2$, or one in the underlying
string theory.  If it is, the symmetry would be novel.  The small
area limit is also interesting because of its implications for
an equation relating the free energy at a given genus to that
at lower genera.  If the
free energy satisfies a string master equation like the
holomorphic anomaly equation, it must satisfy it as $\Atil \ra 0$.
It is very easy to check that no simple equation will work.
Of course, the equation could be more complicated.  For example,
if the full $SU(N)$ partition function satisfies the simple
holomorphic anomaly equation, then the chiral $U(N)$ partition
function will satisfy a non-polynomial (in $1/N$) differential
equation resulting from (\ref{etadiff}).

\section{Conclusions}

In the recent program to extract a string theory from Migdal's
explicit solution of QCD$_2$, we have taken a step backward
from the starting point, in a sense.  We have computed even
more explicit expressions for the QCD$_2$ string free energy
up to genus 8.  These calculations relied on a strong/weak
gauge coupling duality that is exact at each order of string
perturbation theory (but is violated by the non-perturbative
corrections).  The modular structure of the free energy is
familiar from topological string theory, but there does not
seem to be a simple holomorphic anomaly equation for $F^+$.

It might be expected that there would exist a handle generating
operator  since $F^+$ receives no contribution coming from the
boundary of moduli space (collapsed handles or tubes).  In some
sense the differential operator generating the partition function
from the generalized eta function (\ref{etadiff}) plays this
role.  It is not as simple as one would
like, since it couples to each row number separately.  This is
equivalent to having it couple to the infinitely many
deformations of QCD$_2$ (the higher Casimir perturbations),
rather than coupling to the K\"{a}hler modulus $\Atil$ alone.
So it is an important open question in the worldsheet
theory to understand how the free energy at a given genus
is related to that at lower genera.

Another interesting question we have raised is the cause of
the softening of the $\lambda A \ra 0$ singularity.  We proved
the property by looking directly at the small area limit of
the heat kernel expression for the partition function.  It
would be very interesting to have a worldsheet explanation
for this effect (although it might just be accidental).

In any case, the exact expressions for the free energy
offer many possibilities for further investigations.  The goal
of the recent work on two dimensional Yang-Mills
theory is to make progress toward understanding four
dimensional QCD, or at least to learn more about string
theories without spacetime gravity.  We have exhibited
properties of the two dimensional free energy that could
have a bearing on either of these two interesting goals.

\vspace{.4in}

\noindent
Acknowledgements:

I wish to thank
Robbert Dijkgraaf,
Mike Douglas,
David Gross,
Greg Moore and
Ken Intriligator for many interesting discussions.
Also, many thanks to the Aspen Center for Physics, where some
of this work was done.

\appendix
\renewcommand{\theequation}{\Alph{section}.\arabic{equation}}
\section{Appendix: The Free Energy up to Genus 8}

The $U(N)$ free energy on the torus restricted to one chiral sector
is
\be
F_1^+ = - \frac{\epsilon _0}{24}\Atil - \log \eta
\label{fone}
\ee

\be
F_2^+ = - \frac{\gst ^2}{2!\, 90}\left\{\e \ep + \epp \right\}
\label{ftwo}
\ee

\be
F_3^+ = \frac{\gst ^4}{4!\, 54} \left\{ 7 \epb ^3 +3 \eppb ^2 \right\}
\label{fthree}
\ee

\asize{1.5}
\be
\barcl
{\dst F_4^+} & = & {\dst
\frac{\gst ^6}{6!\, 54} \left\{
27 \e \epb ^4 -36 \eb ^2 \epb ^2 \epp -746 \epb ^3 \epp
\right. } \\
& & {\dst \left. ~~~~~~
+ 12 \eb ^3 \eppb ^2 +246 \e \ep \eppb ^2 - 106 \eppb ^3
\right\} }
\eac
\label{ffour}
\ee

\be
\barcl
{\dst F_5^+} & = & {\dst
\frac{\gst ^8}{8!\, 81} \left\{
162 \eb ^4 \epb ^4 + 5940 \eb ^2 \epb ^5 + 85299 \epb ^6
+ 19821 \eppb ^4
\right. } \\ & & {\dst ~~~~~~
- 216 \eb ^5 \epb ^2 \epp - 8352 \eb ^3 \epb ^3 \epp
- 165388 \e \epb ^4 \epp
} \\ & & {\dst ~~~~~~
+72 \eb ^6 \eppb ^2 + 2928 \eb ^4 \ep \eppb ^2
+ 97980 \eb ^2 \epb ^2 \eppb ^2
} \\ & & {\dst \left. ~~~~~~
+ 291334 \epb ^3 \eppb ^2 - 16896 \eb ^3 \eppb ^3
- 116472 \eb \ep \eppb ^3
\right\} }
\eac
\label{ffive}
\ee

\asize{1.2}
\be
\barcl
{\dst F_6^+} & = & {\dst
\frac{\gst ^{10}}{10!\, 27} \left\{
216 \eb ^7 \epb ^4
+ 12420 \eb ^5 \epb ^5
+ 935442 \eb ^3 \epb ^6
\right.  } \\ & & {\dst ~~~~~~
+ 9054978 \e \epb ^7
- 288 \eb ^8 \epb ^2 \epp
- 17136 \eb ^6 \epb ^3 \epp
} \\ & & {\dst ~~~~~~
- 1798164 \eb ^4 \epb ^4 \epp
- 21957340 \eb ^2 \epb ^5 \epp
- 49812944 \epb ^6 \epp
} \\ & & {\dst ~~~~~~
+ 96 \eb ^9 \eppb ^2
+ 5904 \eb ^7 \ep \eppb ^2
+ 1129560 \eb ^5 \epb ^2 \eppb ^2
} \\ & & {\dst ~~~~~~
+ 16998104  \eb ^3 \epb ^3 \eppb ^2
+ 85070724 \eb \epb ^4 \eppb ^2
- 230768 \eb ^6 \eppb ^3
} \\ & & {\dst ~~~~~~
- 4242768 \eb ^4 \ep \eppb ^3
- 43083696 \eb ^2 \epb ^2 \eppb ^3
- 55574424 \epb ^3 \eppb ^3
} \\ & & {\dst \left. ~~~~~~
+ 5750160 \eb ^3 \eppb ^4
+ 22892460 \e \ep \eppb ^4
- 2132916 \eppb ^5
\right\} }
\eac
\label{fsix}
\ee

\be
\barcl
{\dst F_7^+} & = & {\dst
\frac{\gst ^{12}}{12!\, 81} \left\{
2592 \eb ^{10} \epb ^4
 + 204768 \eb ^8 \epb ^5
 + 89976744 \eb ^6 \epb ^6
\right.  } \\ & & {\dst ~~~~~~
 + 1909213524 \eb ^4 \epb ^7
 - 3456 \eb ^{11} \epb ^2 \epp
 - 279936 \eb ^9 \epb ^3 \epp
} \\ & & {\dst ~~~~~~
 - 177698880 \eb ^7 \epb ^4 \epp
 - 4174329312 \eb ^5 \epb ^5 \epp
} \\ & & {\dst ~~~~~~
 - 62394892148 \eb ^3 \epb ^6 \epp
 + 1152 \eb ^{12} \eppb ^2
} \\ & & {\dst ~~~~~~
 + 95616 \eb ^{10} \ep \eppb ^2
 + 116614944 \eb ^8 \epb ^2 \eppb ^2
} \\ & & {\dst ~~~~~~
 + 3013923216 \eb ^6 \epb ^3 \eppb ^2
 + 63439748328 \eb ^4 \epb ^4 \eppb ^2
} \\ & & {\dst ~~~~~~
 - 25422720 \eb ^9 \eppb ^3
 - 719490240 \eb ^7 \ep \eppb ^3
} \\ & & {\dst ~~~~~~
 - 26470959696 \eb ^5 \epb ^2 \eppb ^3
 - 216898783824 \eb ^3 \epb ^3 \eppb ^3
} \\ & & {\dst ~~~~~~
 + 3707598864 \eb ^6 \eppb ^4
 + 44962956000 \eb ^4 \ep \eppb ^4
} \\ & & {\dst ~~~~~~
 - 27342815040 \eb ^3 \eppb ^5
 + 21435613473 \eb ^2 \epb ^8
} \\ & & {\dst ~~~~~~
 + 354642271752 \eb ^2 \epb ^5 \eppb ^2
 + 246384220752 \eb ^2 \epb ^2 \eppb ^4
} \\ & & {\dst ~~~~~~
 - 221060144076 \e \epb ^7 \epp
 - 540228415584 \e \epb ^4 \eppb ^3
} \\ & & {\dst ~~~~~~
 - 75540784608 \e \ep \eppb ^5
 + 37289952912 \epb ^9
} \\ & & {\dst ~~~~~~
 + 343299239380 \epb ^6 \eppb ^2
 + 186490620756 \epb ^3 \eppb ^4
} \\ & & {\dst \left. ~~~~~~~~~~~~~~~~
 + 4465217052 \eppb ^6
\right\} }
\eac
\label{fseven}
\ee

\asize{1.5}
\be
\barcl
{\dst F_8^+} & = & {\dst
\frac{\gst ^{14}}{14!\, 243} \left\{
31104 \eb ^{13} \epb ^4
 + 3146688 \eb ^{11} \epb ^5
 + 9381744000 \eb ^9 \epb ^6
\right.  } \\ & & {\dst ~~~~~~
 + 284491245600 \eb ^7 \epb ^7
 - 41472 \eb ^{14} \epb ^2 \epp
 - 4278528 \eb ^{12} \epb ^3 \epp
} \\ & & {\dst ~~~~~~
 - 27941270346864 \eb ^6 \epb ^6 \epp
 - 606326798592 \eb ^8 \epb ^5 \epp
} \\ & & {\dst ~~~~~~
 - 18712973376 \eb ^{10} \epb ^4 \epp
 + 13824 \eb ^{15} \eppb ^2
 + 1453824 \eb ^{13} \ep  \eppb ^2
} \\ & & {\dst ~~~~~~
 + 12436238592 \eb ^{11} \epb ^2 \eppb ^2
 + 428976125568 \eb ^9 \epb ^3 \eppb ^2
} \\ & & {\dst ~~~~~~
 + 27464636483952 \eb ^7 \epb ^4 \eppb ^2
 - 2753687808 \eb ^{12} \eppb ^3
} \\ & & {\dst ~~~~~~
 - 100795663872 \eb ^{10} \ep  \eppb ^3
 - 11670697945728 \eb ^8 \epb ^2 \eppb ^3
} \\ & & {\dst ~~~~~~
 - 191631253050048 \eb ^6 \epb ^3 \eppb ^3
 + 1796319251712 \eb ^9 \eppb ^4
} \\ & & {\dst ~~~~~~
 + 35809340742912 \eb ^7 \ep  \eppb ^4
 - 65393019413760  \eb ^6 \eppb ^5
} \\ & & {\dst ~~~~~~
 + 10411506055692 \eb ^5 \epb ^8
 + 374290947363144 \eb ^5 \epb ^5 \eppb ^2
} \\ & & {\dst ~~~~~~
 + 622325551111920 \eb ^5 \epb ^2 \eppb ^4
 - 314884801430312 \eb ^4 \epb ^7 \epp
} \\ & & {\dst ~~~~~~
 - 1915319874231576 \eb ^4 \epb ^4 \eppb ^3
 - 599386224758688 \eb ^4 \ep  \eppb ^5
} \\ & & {\dst ~~~~~~
 + 95567800131858 \eb ^3 \epb ^9
 + 2553509520593988 \eb ^3 \epb ^6 \eppb ^2
} \\ & & {\dst ~~~~~~
 + 3327996287536944 \eb ^3 \epb ^3 \eppb ^4
 + 207585191777040 \eb ^3 \eppb ^6
} \\ & & {\dst ~~~~~~
 - 1483200593325666 \eb ^2 \epb ^8 \epp
 - 6326492920454424 \eb ^2 \epb ^5 \eppb ^3
} \\ & & {\dst ~~~~~~
 - 2174062501379952 \eb ^2 \epb ^2 \eppb ^5
 + 287886256181076 \e  \epb ^{10}
} \\ & & {\dst ~~~~~~
 + 4807697809119300 \e  \epb ^7 \eppb ^2
 + 5189221094011812 \e  \epb ^4 \eppb ^4
} \\ & & {\dst ~~~~~~
 + 436733614643256 \e  \ep  \eppb ^6
 - 1172778987525768 \epb ^9 \epp
} \\ & & {\dst ~~~~~~
 - 3523420607226032 \epb ^6 \eppb ^3
 - 1116320466888120 \epb ^3 \eppb ^5
} \\ & & {\dst \left. ~~~~~~~~~~~~~~~~
 - 17870538853512 \eppb ^7
\right\} }
\eac
\label{feight}
\ee

The full $SU(N)$ free energy on the torus is
\be
F_1 = \frac{\epsilon _0}{12}\left( \frac{\lambda A}{2} \right)
- 2 \log \eta
\ee

\be
F_2 = - \frac{\gst ^2}{2!\, 45}\left\{ \e \ep + \epp \right\}
+ \frac{\gsu}{2!\, 6}\ep
\ee

\be
\barcl
{\dst F_3} & = &
{\dst \frac{\gst ^4}{4!\, 27} \left\{ 7 \epb ^3 +3\eppb ^2\right\}
- \frac{\gst ^2 \gsu }{4!\, 3} \left\{
4 \ep \epp \right\}
} \\ & & ~~~~~~~~~~~~~{\dst
+ \frac{\gsu ^2}{4!\, 4} \left\{
4 \e \epp - 5 [\ep ]^2 \right\} }
\eac
\ee

\asize{1.7}
\be
\barcl
{\dst F_4} & = & {\dst
\frac{\gst ^6}{6!\, 27} \left\{
27 \e \epb ^4 -36 \eb ^2 \epb ^2 \epp -746 \epb ^3 \epp
\right. } \\
& & {\dst \left. ~~~~~~~~~~~~~~
+ 12 \eb ^3 \eppb ^2 +246 \e \ep \eppb ^2 - 106 \eppb ^3
\right\} } \\
& & {\dst ~~~~~~
+ \frac{\gst ^4 \gsu }{6!\, 3} \left\{
150 \epb ^4 - 160 \e \epb ^2 \epp
+ 40 \eb ^2 \eppb ^2 + 180 \ep \eppb ^2
\right\} } \\
& & {\dst ~~~~~~
+ \frac{\gst ^2 \gsu ^2}{6!} \left\{
30 \e \epb ^3 - 20 \eb ^2 \ep \epp
-120 \epb ^2 \epp + 60 \e \eppb ^2
\right\} } \\
& & {\dst ~~~~~~
+ \frac{\gsu ^3}{6!\, 24} \left\{
65 \epb ^3 -360 \eb ^2 \epb ^2
+240 \eb ^3 \epp + 420 \e \ep \epp - 480 \eppb ^2
\right\} }
\eac
\ee
The free energy for genus 5, 6, 7 and 8 may be computed
from (\ref{ffive}) to (\ref{feight}), but they are omitted
to save space.

\end{document}